\documentclass[aps,floatfix,onecolumn,notitlepage,superscriptaddress,preprintnumbers,nofootinbib,11pt]{revtex4-2}
\usepackage{natbib,setspace}
\usepackage{graphicx,tikz-feynman}
\usepackage{amssymb}
\usepackage{amsmath,amssymb,stix}
\usepackage{color}
\usepackage{caption}
\usepackage{subcaption}
\usepackage{slashed}
\usepackage{booktabs}

\usepackage{fancyhdr}
\definecolor{darkblue}{rgb}{0,0,0.5}
\usepackage[colorlinks,linkcolor=darkblue,citecolor=darkblue,urlcolor=darkblue]{hyperref}

\usepackage[capitalize]{cleveref}
\allowdisplaybreaks

\newcommand\beq{\begin{equation}}
\newcommand\eeq{\end{equation}}

\begin{document}

%%%%%%%%%%%%%%%%%%%%%%%%%%%%%%%%%%%%%%%%%%%%%%%%%%%%%%%%%%%%%%%%%%

\title{Deep Neural Networks for Heavy Lepton-Flavor-Violating Higgs Searches at the LHC}

%\title{The 146 GeV $H\to e\mu$ excess}
%%%%%%%%%%%%%%%%%%%%%%%%%%%%%%%%%%%%%%%%%%%%%%%%%%%%%%%%%%%%%%%%%%

\author{Akmal Ferdiyan}
\email{aferdiyan@unsoed.ac.id}
\affiliation{Theoretical High Energy Physics Research Division,
Faculty of Mathematics and Natural Sciences, Institut Teknologi Bandung,
Jl. Ganesha no.10 Bandung, 40132, Indonesia }
\author{Reinard Primulando}
\email{rprimulando@unpar.ac.id}
\affiliation{Center for Theoretical Physics, Department of Physics, Parahyangan Catholic University, Jalan Ciumbuleuit 94, Bandung 40141, Indonesia}
\author{Fiki Taufik Akbar}
\email{ftakbar@itb.ac.id}
\affiliation{Theoretical High Energy Physics Research Division,
Faculty of Mathematics and Natural Sciences, Institut Teknologi Bandung,
Jl. Ganesha no.10 Bandung, 40132, Indonesia }
\author{Bobby Eka Gunara}
\email{bobby@itb.ac.id}
\affiliation{Theoretical High Energy Physics Research Division,
Faculty of Mathematics and Natural Sciences, Institut Teknologi Bandung,
Jl. Ganesha no.10 Bandung, 40132, Indonesia }

%%%%%%%%%%%%%%%%%%%%%%%%%%%%%%%%%%%%%%%%%%%%%%%%%%%%%%%%%%%%%%%%%%
\begin{abstract}
We study lepton-flavor-violating (LFV) decays of a heavy Higgs boson, 
$H \to \mu\tau$, in the Type-III two-Higgs-doublet model by recasting 
the CMS search at $\sqrt{s} = 13$~TeV with 35.9~fb$^{-1}$ using fast 
detector simulation in the mass range 200--450~GeV. We develop a deep 
neural network (DNN) classifier trained on final-state kinematic 
variables that, with mass-dependent threshold optimization, reduces the 
expected 95\% CL upper limits on the signal cross section by 42--46\% 
in the 0-jet channel and 36--40\% in the 1-jet channel relative to the 
standard collinear mass ($M_\mathrm{col}$) baseline. We apply SHAP 
interpretability analysis to identify the visible mass $m_\mathrm{vis}$ 
as one of the dominant discriminating feature, reflecting the characteristic 
neutrino momentum fraction of the $\tau$ decay. We show that 
supplementing the $M_\mathrm{col}$ analysis with a simplified 
mass-dependent pre-selection, $m_\mathrm{vis} < f \cdot m_H$ with 
$f = 0.7$ (0-jet) and $f = 0.8$ (1-jet), consistently improves the 
sensitivity over the $M_\mathrm{col}$-only baseline without requiring 
multivariate infrastructure. In addition, a DNN regression model trained 
to predict the ratio $m_H/M_\mathrm{col}$ corrects the systematic 
prediction bias inherent in the collinear approximation, maintaining an 
absolute mass prediction error below 1~GeV for signals up to 400~GeV 
and improving the mass resolution by 12\% (0-jet) and 21\% (1-jet) at 
$m_H = 450$~GeV. These results demonstrate a clear path toward 
significantly enhanced sensitivity in LFV Higgs searches at the LHC.
\end{abstract}
%%%%%%%%%%%%%%%%%%%%%%%%%%%%%%%%%%%%%%%%%%%%%%%%%%%%%%%%%%%%%%%%%%

\maketitle
\newpage

\flushbottom
%%%%%%%%%%%%%%%%%%%%%%%%%%%%%%%%%%%%%%%%%
\section{Introduction}
%%%%%%%%%%%%%%%%%%%%%%%%%%%%%%%%%%%%%%%%

The discovery of the 125~GeV Higgs boson at the Large Hadron Collider 
(LHC)~\cite{ATLAS:2012gk, CMS:2012qbp} has confirmed the Standard Model 
(SM) mechanism of electroweak symmetry breaking. The SM, however, remains 
an incomplete description of nature, failing to account for dark matter, 
neutrino masses, and the baryon asymmetry of the universe. These 
shortcomings motivate extensions of the SM, many of which predict lepton 
flavor violation (LFV) in the Higgs sector. Such LFV interactions, though 
absent in the SM, arise naturally in two-Higgs-doublet 
models~\cite{Branco:2011iw, Lee:1973iz}, supersymmetric 
models~\cite{Arhrib:2012ax, Goudelis:2011un, Arganda:2005ji}, composite 
Higgs models~\cite{Agashe:2009di}, models with warped extra 
dimensions~\cite{Azatov:2009na, Buras:2009ka}, and 
others~\cite{Harnik:2012pb, Omura:2015nja}. In particular, the Type-III 
two-Higgs-doublet model (2HDM) provides a minimal and well-motivated 
framework for generating lepton flavor violation at tree level through the 
structure of its Yukawa sector~\cite{Primulando:2016eod}. Any observation 
of LFV Higgs decays would constitute unambiguous evidence of physics 
beyond the Standard Model.

Searches for LFV decays of the Higgs boson have been carried out by the 
ATLAS and CMS collaborations. For the 125~GeV Higgs boson, both 
collaborations have established limits at the $\mathcal{O}(0.1\%)$ level 
on the branching fractions $\mathcal{B}(h \to \tau\mu)$ and 
$\mathcal{B}(h \to \tau e)$~\cite{ATLAS:2023mvd, CMS:2021rsq, 
CMS:2015qee, ATLAS:2016jfq}, while the constraint on 
$\mathcal{B}(h \to \mu e)$ is considerably more stringent, at the 
$\mathcal{O}(10^{-5})$ level~\cite{CMS:2023pte, ATLAS:2019old}. The 
search for heavier LFV resonances provides complementary sensitivity and, 
in the framework of the Type-III 2HDM, can probe a substantially larger 
region of the parameter space~\cite{Primulando:2016eod}. The CMS collaboration has conducted searches for 
heavy Higgs decays to $\tau\mu$ and $\tau e$ in the 200--900~GeV mass 
range~\cite{CMS:2019pex}, as well as for heavy resonances decaying to 
$\mu e$~\cite{CMS:2023pte}, with the latter reporting a hint of a signal 
at approximately 146~GeV. No conclusive evidence for LFV Higgs decays 
has been found in any of these searches.

The existing heavy LFV Higgs searches, involving tau leptons, rely on the collinear mass 
approximation $M_\mathrm{col}$ as the primary discriminant. This 
approximation assumes that the neutrinos from the $\tau$ decay are emitted 
collinear with the visible decay products, and therefore does not exploit 
the full kinematic information of the final state. %Moreover, the collinear  assumption becomes increasingly inaccurate for boosted $\tau$ leptons at  high mass, introducing a systematic bias that grows with $m_H$. 
Deep  neural networks (DNNs) have emerged as powerful tools in the analysis of 
LHC data~\cite{Baldi:2016fzo,Guest:2018yhq, Albertsson:2018maf,Feickert:2021ajf}, and have been applied 
to improve the sensitivity of SM Higgs searches in tau final 
states~\cite{Baldi:2014pta,Bartschi:2019xlg, CMS:2022kdi}. 
However, no attempt has been made to apply these techniques to the search 
for LFV Higgs decays.

In this paper, we recast the CMS search for heavy LFV Higgs 
bosons~\cite{CMS:2019pex} and propose an enhanced analysis framework 
based on deep learning. We develop a DNN classifier trained on the 
final-state kinematic variables that achieves a 36--46\% reduction in 
the expected 95\% CL upper limits on the signal cross section relative 
to the $M_\mathrm{col}$ baseline. In order to interpret the classifier's 
decision, we employ SHAP (SHapley Additive 
exPlanations)~\cite{Lundberg:2017uca} values, which identify the visible 
mass $m_\mathrm{vis}$ as one of the dominant discriminating features and lead to 
a simplified mass-dependent cut that captures a significant fraction of 
the DNN's sensitivity gain without requiring multivariate infrastructure. 
In addition, we develop a DNN regression model that corrects the 
systematic prediction error inherent in the collinear approximation, 
maintaining an absolute mass prediction error below 1~GeV for signals 
up to 400~GeV.

The paper is organized as follows. In Sec.~\ref{sec:model} we give a 
brief overview of the Type-III 2HDM. In Sec.~\ref{sec:htaumu_search} we 
describe the signal and background processes, define the key kinematic 
variables, and detail the event selection and simulation. The DNN 
classifier, its interpretability analysis, and the resulting sensitivity 
limits are presented in Sec.~\ref{sec:NeurNet}. The mass regression study 
is discussed in Sec.~\ref{sec:masspredict}. We conclude in 
Sec.~\ref{sec:conc}.

%%%%%%%%%%%%%%%%%%%%%%%%%%%%%%%%%%%%%%%%%
\section{Type-III Two-Higgs-Doublet Model}
\label{sec:model}

Extending the scalar sector of the Standard Model by a second $SU(2)_L$ 
doublet provides a minimal and well-motivated framework for generating 
lepton flavor violation at tree level. Unlike the Type-I and Type-II 
variants, the Type-III two-Higgs-doublet model (2HDM) imposes no discrete 
symmetry to separate the Yukawa couplings of the two doublets from the 
fermion sector. As a consequence, both doublets couple simultaneously to 
all fermion species, and the resulting Yukawa matrices are generically 
non-diagonal in flavor space~\cite{Primulando:2016eod}.

A particularly transparent parametrization of the model is afforded by 
the Higgs basis~\cite{Georgi:1978fu}. This particular basis had the vacuum structure made manifest by rotating the two doublets, both carrying hypercharge 
$Y = 1/2$, into a frame where only one of them ($\Phi_1$) acquires a 
nonzero vacuum expectation value $v$. The second doublet $\Phi_2$ is 
inert with respect to electroweak symmetry breaking. In unitary gauge, 
the two doublets take the form:
\begin{equation}
\Phi_1 =
\begin{pmatrix}
0 \\
\dfrac{1}{\sqrt{2}}(v + \phi_1)
\end{pmatrix},
\qquad
\Phi_2 =
\begin{pmatrix}
H^+ \\
\dfrac{1}{\sqrt{2}}(\phi_2 + iA)
\end{pmatrix}.
\end{equation}

The physical mass eigenstates are obtained by diagonalizing the 
CP-even neutral sector through a rotation by mixing angle $\alpha$:
\begin{equation}
\begin{pmatrix}
\phi_1 \\
\phi_2
\end{pmatrix}
=
\begin{pmatrix}
c_\alpha & s_\alpha \\
-s_\alpha & c_\alpha
\end{pmatrix}
\begin{pmatrix}
h \\
H
\end{pmatrix},
\end{equation}
where $c_\alpha \equiv \cos\alpha$ and $s_\alpha \equiv \sin\alpha$. 
The resulting physical spectrum consists of two CP-even neutral scalars 
$h$ and $H$ (with $m_h < m_H$), one CP-odd neutral scalar $A$, and a 
charged scalar pair $H^\pm$. The lighter state $h$ is identified with 
the 125~GeV Higgs boson observed at the LHC, while the heavier state 
$H$ is the new scalar whose LFV decays are studied in this work.

The Yukawa sector of the model is the key source of phenomenological 
interest. Fermion masses are generated by $\Phi_1$ through the standard 
Higgs mechanism, while $\Phi_2$ introduces additional coupling matrices 
$Y_f$ that need not respect flavor diagonality. The full Yukawa Lagrangian 
reads~\cite{Primulando:2016eod}:
\begin{align}
\mathcal{L}_{\text{yuk}} =
&- \bar{L}_L \frac{\sqrt{2} m_\ell}{v} \, \ell_R \Phi_1
- \sqrt{2} \bar{L}_L Y_\ell \ell_R \Phi_2
\nonumber \\
&- \bar{Q}_L \frac{\sqrt{2} m_U}{v} \, u_R \tilde{\Phi}_1
- \sqrt{2} \bar{Q}_L V Y_u u_R \tilde{\Phi}_2
\nonumber \\
&- \bar{Q}_L V \frac{\sqrt{2} m_D}{v} \, d_R \Phi_1
- \sqrt{2} \bar{Q}_L V Y_d d_R \Phi_2,
\end{align}
where $m_f$ are the diagonal fermion mass matrices, $Y_f$ are the 
generally non-diagonal Yukawa coupling matrices associated with $\Phi_2$, 
$\tilde{\Phi} = i\sigma_2\Phi^*$ is the conjugate doublet, and $V$ denotes 
the Cabibbo--Kobayashi--Maskawa (CKM) matrix. All fermion fields are 
expressed in the mass eigenbasis, with $SU(2)_L$ doublet assignments:
\begin{equation}
L_L =
\begin{pmatrix}
\nu_L \\
\ell_L
\end{pmatrix},
\qquad
Q_L =
\begin{pmatrix}
u_L \\
V d_L
\end{pmatrix}.
\end{equation}
It is precisely the off-diagonal entries of $Y_\ell$ that mediate 
flavor-changing transitions between charged leptons via the exchange 
of $H$, $A$, or $H^\pm$. In the lepton sector, the coupling of the 
heavy CP-even scalar $H$ to a $\mu\tau$ pair is proportional to 
$(Y_\ell)_{\mu\tau}$ and $(Y_\ell)_{\tau\mu}$, making the decay 
$H\to\mu\tau$ a direct and clean probe of the off-diagonal Yukawa 
texture. This decay is the focus of the present analysis.

\section{$H \rightarrow \mu \tau$  search strategy}
\label{sec:htaumu_search}

\subsection{Signal and Backgrounds}
\label{sec:signalandbakcgrounds}
%\label{sec:htaumu_search}
%\subsection{Events Selection}

We consider the production of $H$ at the LHC via gluon fusion, followed by 
the LFV decay $H \to \mu\tau$. The tau lepton subsequently decays either 
leptonically or hadronically. This study focuses on the leptonic decay mode, $\tau^- \to e^-\bar{\nu}_e\nu_\tau$, resulting in an opposite-sign pair $e \mu$ and missing transverse energy ($E_T^\mathrm{miss}$) in the final state. This channel provides a cleaner experimental signature than the hadronic mode,  which requires dedicated $\tau_h$ identification algorithms to separate genuine  hadronic taus from misidentified quark and gluon jets. The jet$\to\tau_h$ fake rate and the kinematic properties of the resulting 
fake candidates are not reliably reproduced by fast simulation frameworks 
such as \textsc{Delphes}~\cite{deFavereau:2013fsa}. This is because genuine 
hadronic taus have a distinctive collimated, low-multiplicity structure that 
the real $\tau_h$ identification algorithms exploit, while \textsc{Delphes} 
relies on simplified parametric efficiencies that do not capture this 
substructure. For these reasons, we restrict the present analysis  to the leptonic $\tau$ decay mode.

The dominant SM background processes contributing to the $e\mu + E_T^\mathrm{miss}$ final state are $W^+W^-$ diboson production with leptonic decays of both $W$ bosons, and $t\bar{t}$ production with semi-leptonic decays of both top quarks. The cross sections of additional diboson processes ($WZ$, $ZZ$) are smaller by an order of magnitude, and are therefore not included in the present analysis. Other backgrounds such as the $Z\rightarrow\tau \tau$ process, while significant at lower mass scales, this process is suppressed in the search region ($m_H \geq 200$ GeV) the rapidly falling Drell-Yan tail.

%

%In this work, we focus on the production of the Higgs boson in proton--proton collisions followed by the LFV decay $pp \rightarrow H \rightarrow  \mu \tau$. The tau lepton subsequently decays either leptonically or hadronically. Due to the presence of neutrinos in the tau decay, the final state contains missing transverse momentum in addition to the visible decay products. The resulting event topology therefore consists of a muon, the visible decay products of the tau lepton, and missing transverse energy.

\subsection{Event selections}
\label{sec:eventselect}
Our work follows the CMS search for $H \rightarrow \mu\tau$~\cite{CMS:2019pex}, which also serves as the benchmark. Specifically, we investigate the $H \rightarrow \mu \tau_{e}$ decay channel. The search utilizes 35.9 fb$^{-1}$ of the LHC run data with the corresponding $\sqrt{s} = 13$ TeV. The CMS search requires the events to have two oppositely charged light leptons with different flavors and separated by $\Delta R > 0.3$. The electrons must satisfy the requirements of $p_T>10$ GeV and $|\eta|<2.4$; however, events with additional electron candidates with $p_T > 5$ GeV are rejected. The muons are required to have $p_T>53$ GeV and $|\eta|<2.4$, and events with additional $\mu$ candidates with $p_T>10$ GeV are likewise rejected. The events are subsequently divided into 0-jet and 1-jet categories. The jets must meet the criteria of $p_T>30$ GeV and $|\eta|<4.7$. Events with more than one jet are discarded. % The collinear mass, $M_{col}$, is used to approximates the Higgs mass in place of visible mass $m_\mathrm{vis}$, given that $M_{col}$ offers better background-signal separation.

Two mass variables are used extensively in this analysis. The visible 
mass $m_\mathrm{vis} = \sqrt{(p_\mu + p_e)^2}$ is the invariant mass of 
the muon and electron reconstructed directly from their four-momenta, 
and systematically underestimates $m_H$ since it ignores the neutrino 
contributions. The collinear mass $M_\mathrm{col} = m_\mathrm{vis}/\sqrt{x_\tau}$, 
where $x_\tau = |\vec{p}_T^{\,e}|/(|\vec{p}_T^{\,e}| + |\vec{p}_T^{\,\rm miss}|)$ 
is the fraction of the tau momentum carried by the visible decay products, 
improves upon this by assuming the neutrinos are emitted collinear with 
the visible tau decay products. While $M_\mathrm{col}$ provides a better 
estimate of $m_H$ than $m_\mathrm{vis}$ in most events, it carries persistent systematic prediction bias arising from the propagation of $E_T^\mathrm{miss}$ reconstruction uncertainties through the mass formula. 

%The goal of this analysis is to distinguish the LFV Higgs signal from the dominant Standard Model background processes using kinematic information reconstructed from the final-state particles.

\subsection{Simulations and Validations}
Signal events for the process $pp\to H\to\mu\tau$ with $\tau\to
e\nu_e\nu_\tau$ are generated at leading order using
\textsc{MadGraph5}~\cite{Alwall:2014hca} with the SM particle content
supplemented by a flavor-violating Yukawa interaction in the Type-III 2HDM
framework.
Parton showering and hadronization are
performed with \textsc{Pythia~8}~\cite{Sjostrand:2014zea}. A fast detector simulation is carried out with
\textsc{Delphes~3}~\cite{deFavereau:2013fsa} employing the CMS detector card.
Further analysis, including the extraction of kinematic variables, is performed
within the \textsc{MadAnalysis~5} framework~\cite{Conte:2012fm,Conte:2014zja}. To simulate $\tau$ decay, we also used the TauDecay library \cite{Hagiwara:2012vz} inside the MadGraph framework, to subsequently decay the $\tau$ to $e, \nu_e$ and $\nu_\tau$. The TauDecay package handles the tau decay at the matrix-element level, preserving the spin correlations between the production and decay stages. This ensures that the kinematic features fed to both the classifier and the regression network correctly reflect the polarization state of the tau, which is essential for the validity of the analysis.

\begin{figure}
     \centering
     \begin{subfigure}[b]{0.48\textwidth}
         \centering
         \includegraphics[width=\textwidth]{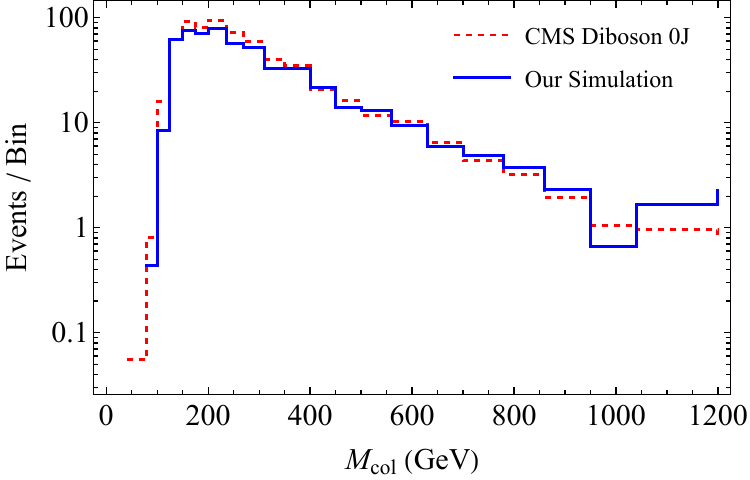}
         \caption{Diboson background}
         \label{fig:Diboson0j}
     \end{subfigure}
     \hfill
     \begin{subfigure}[b]{0.48\textwidth}
         \centering
         \includegraphics[width=\textwidth]{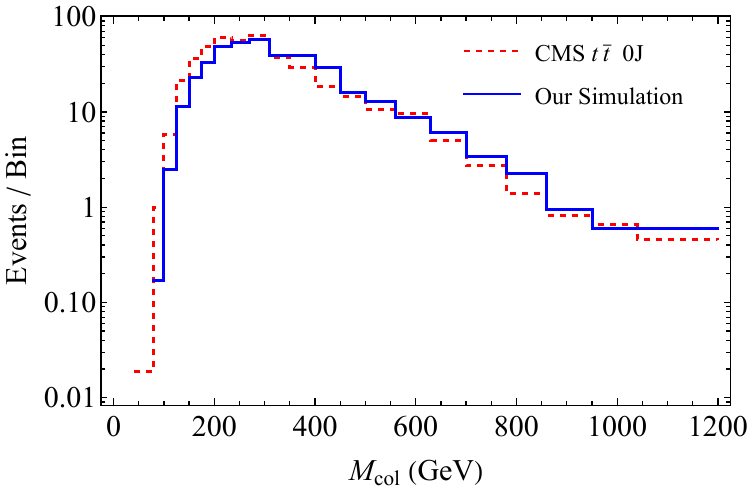}
         \caption{$t \bar{t}$ background}
         \label{fig:ttbar0j}
     \end{subfigure}
\caption{Validation of simulated background distributions for the 0-jet channel. The distributions of $M_\mathrm{col}$ for (a) diboson and (b) $t\bar{t}$ processes are compared against the CMS results~\cite{CMS:2019pex}.}
        \label{fig:comparison0j}
\end{figure}

\begin{figure}
     \centering
     \begin{subfigure}[b]{0.48\textwidth}
         \centering
         \includegraphics[width=\textwidth]{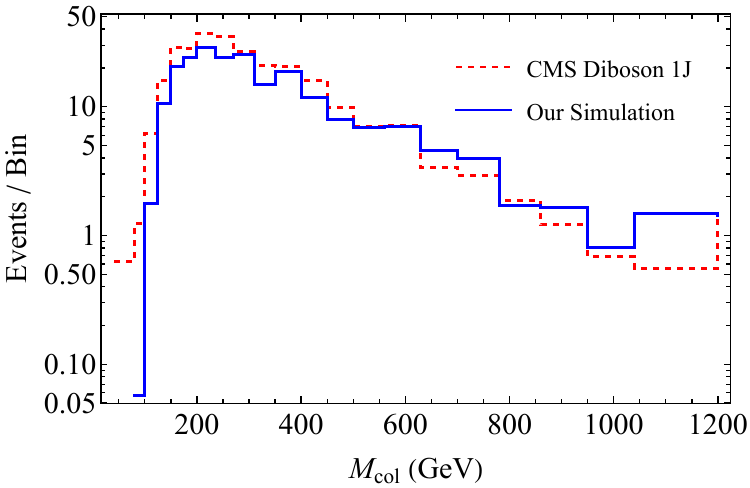}
         \caption{Diboson background}
         \label{fig:Diboson1j}
     \end{subfigure}
     \hfill
     \begin{subfigure}[b]{0.48\textwidth}
         \centering
         \includegraphics[width=\textwidth]{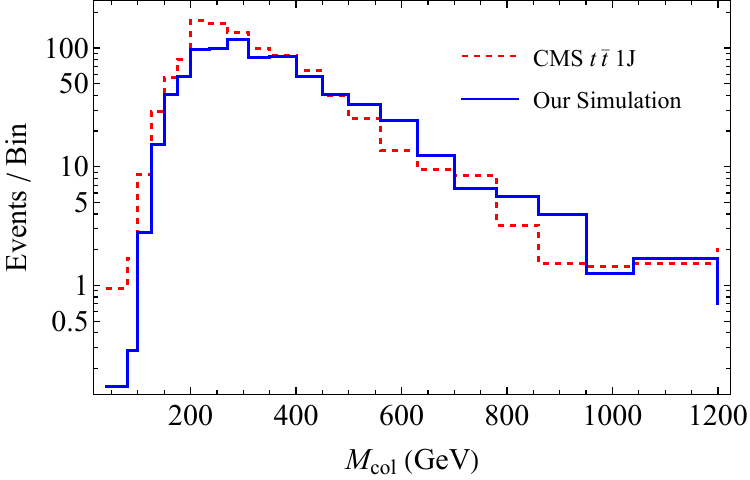}
         \caption{$t \bar{t}$ background}
         \label{fig:ttbar1j}
     \end{subfigure}
\caption{Validation of simulated background distributions for the 1-jet channel. The distributions of $M_\mathrm{col}$ for (a) diboson and (b) $t\bar{t}$ processes are compared against the CMS 13 TeV results~\cite{CMS:2019pex}.}
        \label{fig:comparison1j}
\end{figure}

\begin{figure}
     \centering
     \begin{subfigure}[b]{0.48\textwidth}
         \centering
         \includegraphics[width=\textwidth]{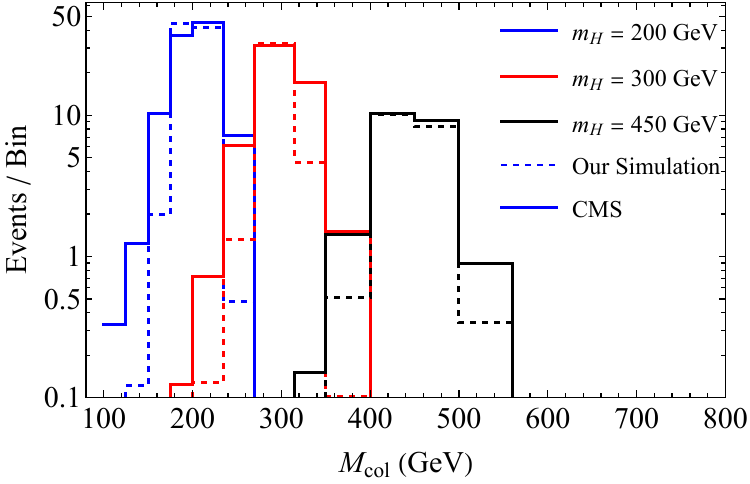}
         \caption{Signal simulation 0-jet}
         \label{fig:signal0j}
     \end{subfigure}
     \hfill
     \begin{subfigure}[b]{0.48\textwidth}
         \centering
         \includegraphics[width=\textwidth]{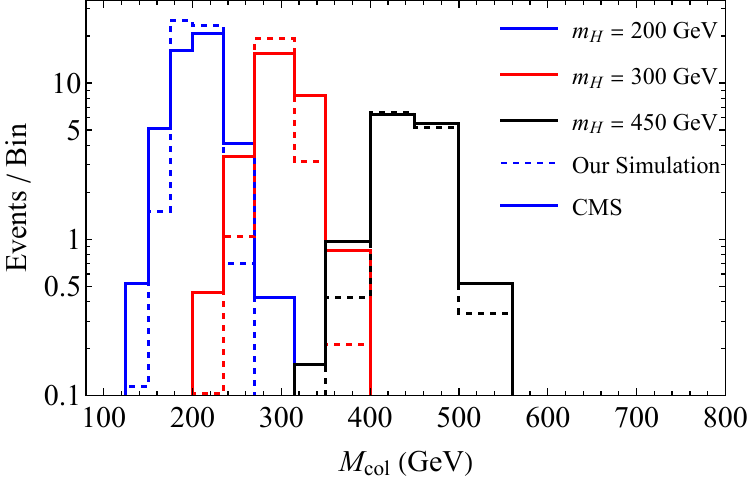}
         \caption{Signal simulation 1-jet}
         \label{fig:signal1j}
     \end{subfigure}
\caption{Signal simulation validation for $m_H = 200, 300$ and $450$ GeV. The  $M_\mathrm{col}$ distributions for the 0-jet (a) and 1-jet (b) channels are compared to CMS signal templates~\cite{CMS:2019pex}.}
        \label{fig:signalcomparison}
\end{figure}

To ensure the reliability of our recasting procedure, we performed a rigorous validation of our simulated templates against the official CMS benchmark results. As illustrated in Figs.~\ref{fig:comparison0j} and ~\ref{fig:comparison1j}, the kinematic distributions for the dominant diboson and $t\bar{t}$ backgrounds show close alignment to the CMS predictions across both the 0-jet and 1-jet categories. Similarly, the signal validation for $m_H = 200, 300$ and $450$ GeV as shown in Fig. ~\ref{fig:signalcomparison} demonstrates that our simulation accurately captures the expected mass resolution and peak positions of the $M_\mathrm{col}$ variable. This close agreement across all channels establishes a robust foundation for the subsequent DNN-based multivariate analysis and limit-setting procedures.

%Background samples for DNN training are generated by running \textsc{MadGraph5\_aMC@NLO} in 100 independent batches of $10^4$ events each, with identical process definitions, run cards, and \textsc{Pythia~8} and \textsc{Delphes~3} configurations, yielding a total of $10^6$ background events per process. After applying the selection criteria described in Sec.\ref{sec:eventselect}, 45,900 and 36,600 events remained in the 0-jet and 1-jet category, respectively. This represents an efficiency of approximately 4.6\% and 3.6\% for each category. 

%The effective cross sections after all selection requirements are obtained from the
%\textsc{MadGraph5} integrated weight, verified to be stable across
%all 100 batches with run-to-run variations below 1\,\%. The effective cross
%sections are $\sigma_{WW}^{\rm eff} = 0.152$~pb and
%$\sigma_{t\bar{t}}^{\rm eff} = 0.087$~pb, implicitly incorporating the
%relevant branching fractions and detector-level selection efficiencies.

To provide the DNN with a continuous and dense representation of the signal across the parameter space, signal events were generated in a fine-grained mass scan from 200 GeV to 450 GeV. We sampled 2,000 discrete mass points with a step size of approximately 0.125 GeV. For each mass hypothesis, $1000$ events were generated in two independent batches of 500, resulting in a total signal training set of $2\times10^6$ events. To construct a balanced training dataset and avoid class-imbalance bias, we performed stratified sampling on this signal pool to match the number of background events passing all selection cuts, for both 0-jet and 1-jet categories. Signal events were selected using 50 equal-width bins in the Higgs mass $m_H$, ensuring a uniform representation of the signal across the entire 200--450~GeV range. This high-density, uniform sampling ensures that the classifier can smoothly interpolate between mass hypotheses and avoids local over-training, while providing the large, quasi-continuous dataset required for the mass regression model described in Sec.~\ref{sec:masspredict}. 

\section{Deep Neural Network Signal/Background Classifier}
\label{sec:NeurNet}

\subsection{DNN architecture and training}
Event-level discrimination is performed using ten kinematic features reconstructed from the visible decay products and missing transverse momentum. The input 
features are the three-momentum components of the muon 
$(p_x^\mu, p_y^\mu, p_z^\mu)$ and electron $(p_x^e, p_y^e, p_z^e)$, 
the transverse components of the missing momentum $(p_x^\nu, p_y^\nu)$, 
the visible mass $m_\mathrm{vis}$, and the collinear mass $M_\mathrm{col}$, 
giving a total of ten input features. All features are standardised to 
zero mean and unit variance prior to training.

The classifier is implemented in \textsc{TensorFlow}~\cite{tensorflow2015-whitepaper} and trained to distinguish signal events from SM background using binary cross-entropy. The architecture consists of five hidden layers to refine feature representations. In the initial high-capacity layers, Batch Normalization (BN) and Dropout ($p=0.3$) are applied to stabilize training and reduce neuron co-dependency. To ensure consistent gradient flow across the full dynamic range of the input features, LeakyReLU ($\alpha=0.1$) is utilized as the activation function throughout all hidden layers. A transitional 32-unit layer is implemented without Batch Normalization, with the purpose of allowing the network to preserve raw activation magnitudes and refine the feature space before the final classification stage. The network's final layer is a Sigmoid output layer, yielding a score $p \in [0,1]$ where values near unity indicate signal-like events. 

Optimization is performed using the Adam algorithm, with a learning rate scheduler and an early stopping criterion (100 epoch patience) ensuring convergence, typically reached within 350 epochs. The characteristic offset between training and validation metrics is a known consequence of the weight-scaling inference rule in Dropout~\cite{JMLR:v15:srivastava14a} and the shift from mini-batch to population statistics in Batch Normalization~\cite{pmlr-v37-ioffe15}. These mechanisms reduce stochastic noise during evaluation, often resulting in validation losses lower than training losses without compromising the model’s generalization capabilities.

The full architecture and 
training configuration are summarised in Tables~\ref{tab:architecture} 
and~\ref{tab:training}. Separate models are trained for the 0-jet and 
1-jet categories.

\begin{table}[h]
\centering
\caption{DNN architecture for signal/background classification.}
\label{tab:architecture}
\begin{tabular}{lcccc}
\hline\hline
Layer & Units & Activation & Dropout & BN \\
\hline
Input       & 10  & —                     & —   & —   \\
Dense       & 128 & LeakyReLU($\alpha=0.1$) & 0.3 & Yes \\
Dense       & 64  & LeakyReLU($\alpha=0.1$) & 0.3 & Yes \\
Dense       & 64  & LeakyReLU($\alpha=0.1$) & —   & Yes \\
Dense       & 32  & LeakyReLU                    & —   & No  \\
Dense       & 32  & LeakyReLU($\alpha=0.1$) & —   & Yes \\
Output      & 1   & Sigmoid               & —   & —   \\
\hline\hline
\end{tabular}
\end{table}

\begin{table}[h]
\centering
\caption{Training configuration for the signal/background classifier.}
\label{tab:training}
\begin{tabular}{ll}
\hline\hline
Hyperparameter    & Value \\
\hline
Optimiser         & Adam \\
Learning rate     & $10^{-3}$, min $10^{-6}$ \\
LR scheduler      & ReduceLROnPlateau (factor 0.5, patience 10) \\
Loss function     & Binary cross-entropy \\
Batch size        & 64 \\
Early stopping    & patience 100, monitor val.\ accuracy \\
Train/test split  & 80\% / 20\% (stratified) \\
Feature scaling   & StandardScaler (zero mean, unit variance) \\
Input features    & $p_x^{\mu}, p_y^{\mu}, p_z^{\mu}, p_x^{e}, p_y^{e}, 
                    p_z^{e}, p_x^{\nu}, p_y^{\nu}, m_\mathrm{vis}, M_\mathrm{col}$ \\
\hline\hline
\end{tabular}
\end{table}

To ensure an unbiased evaluation of the DNN performance and the resulting expected limits, the training and testing phases utilize statistically independent datasets. A separate "test-bench" dataset was generated specifically for the sensitivity analysis, consisting of $10^6$ events for each primary background process ($WW$ and $t\bar{t}$) and $10^5$ events for each signal mass hypothesis. These test samples were produced using independent random seeds and were not seen by the networks during any stage of the training or hyperparameter optimization process. This procedure ensures that the reported results are free from overtraining effects and represent the true generalization power of the multivariate models.

We evaluate the expected 95\,\% CL upper limits on the signal cross-section using two distinct strategies. First, we utilize the DNN score $p \in [0,1]$ as the primary discriminant. Rather than adopting a universal threshold, the cut value $p_{\rm cut}$ is optimized independently for each mass hypothesis by scanning the range [0.1,0.99]. For each threshold, surviving events are binned in $M_\mathrm{col}$ to calculate the profile likelihood test statistic (Eq.~\ref{eq:likelihood}),
\begin{equation}
    q = 2\sum_{i} \left[ s_i - b_i \ln\!\left(1 + \frac{s_i}{b_i}\right) \right],
    \label{eq:likelihood}
\end{equation}
where $s_i$ and $b_i$ are the expected signal and background yields in bin $i$, normalised to an integrated luminosity of $\mathcal{L} = 35.9\ \mathrm{fb}^{-1}$. The 95\% CL limit is obtained by solving $q = 3.84$, corresponding to the one-sided $\chi^2$ critical value for one degree of freedom~\cite{Cowan:2010js}. The value of $p_{\rm cut}$ minimizing the expected limit is adopted as the optimal working point for each mass hypothesis. In this phenomenological study, we focus on the statistical power of the DNN classifier; therefore, systematic uncertainties on the background normalization and shape are not included in the likelihood fit. The resulting limits represent the maximal potential sensitivity under the fast-simulation framework.

\begin{figure}
     \centering
     \begin{subfigure}[b]{0.48\textwidth}
         \centering
         \includegraphics[width=\textwidth]{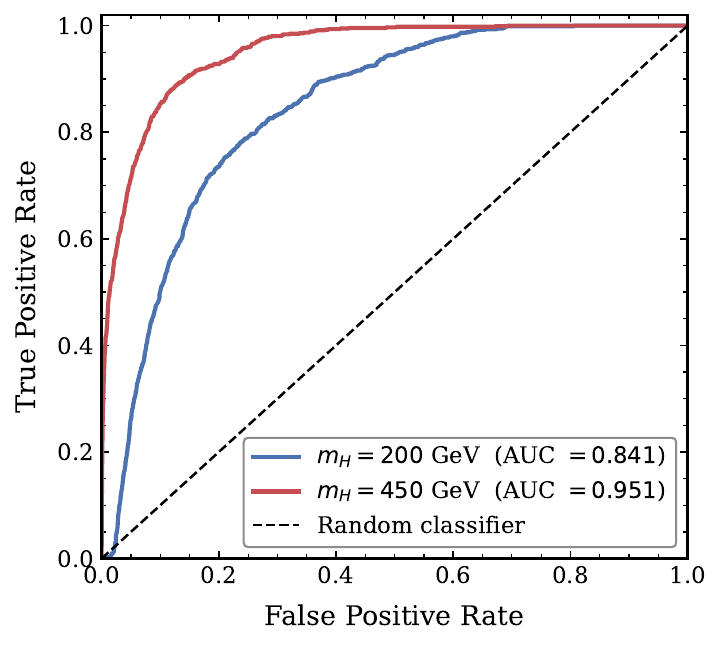}
         \caption{DNN ROC and AUC for 0-jet}
         \label{fig:AUC0j}
     \end{subfigure}
     \hfill
     \begin{subfigure}[b]{0.48\textwidth}
         \centering
         \includegraphics[width=\textwidth]{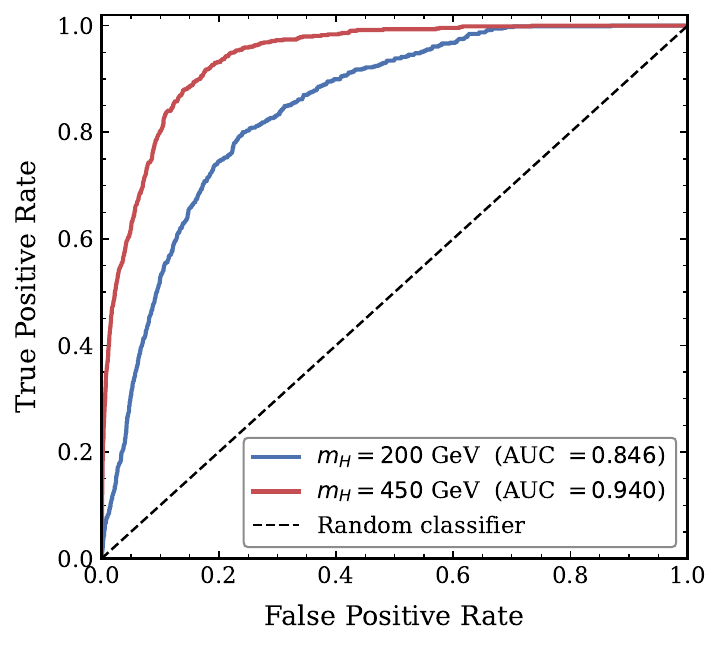}
         \caption{DNN ROC and AUC for 1-jet}
         \label{fig:AUC1j}
     \end{subfigure}
\caption{ROC curves and corresponding AUC values for  $m_H = 200$ and $450$ GeV, for 0-jet and 1-jet categories. The dashed diagonal line represents the performance of a random classifier (AUC = 0.5). The AUC values demonstrate the increased classification power of the network as the signal mass increases.}
        \label{fig:AUC}
\end{figure}

To evaluate the model's performance as a binary classifier, we utilize the Receiver Operating Characteristic (ROC) and Area Under the ROC Curve (AUC), which provides a threshold-independent measure of the network's ability to discriminate between signal and background. An AUC of 1.0 represents perfect separation, while a value of 0.5 indicates performance no better than random chance. The DNN classifier achieves AUC values of 0.841 (0.846) for 
$m_H = 200$~GeV and 0.951 (0.940) for $m_H = 450$~GeV in the 0-jet 
(1-jet) channel, as can be seen in Fig.~\ref{fig:AUC}. 

\begin{figure}
     \centering
     \begin{subfigure}[b]{0.48\textwidth}
         \centering
         \includegraphics[width=\textwidth]{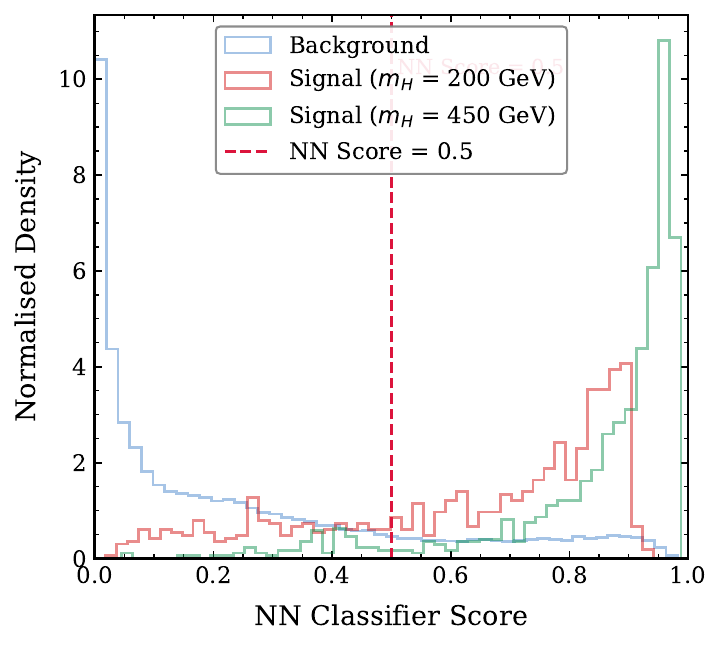}
         \caption{DNN classifier score for 0-jet channel}
         \label{fig:nnclass0j}
     \end{subfigure}
     \hfill
     \begin{subfigure}[b]{0.48\textwidth}
         \centering
         \includegraphics[width=\textwidth]{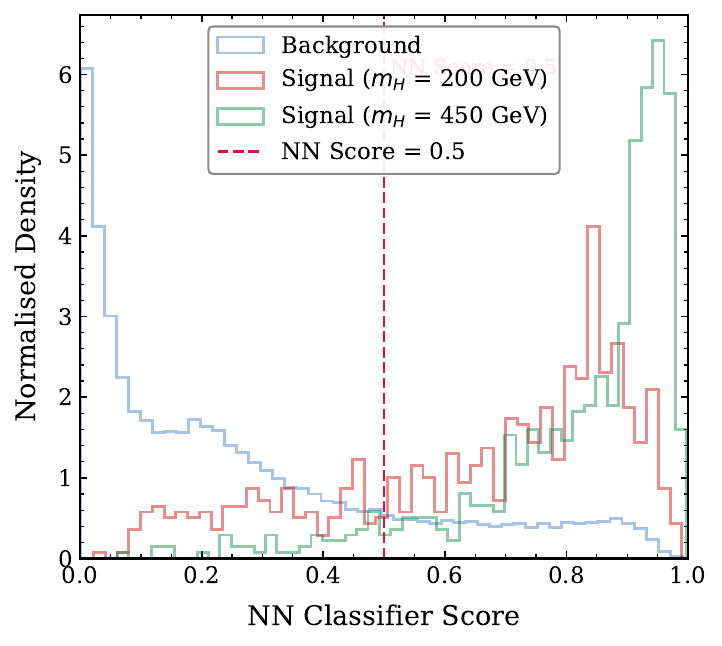}
         \caption{DNN classifier score for 1-jet channel}
         \label{fig:nnclass1j}
     \end{subfigure}
\caption{DNN classifier score distributions for (a) 0-jet and (b) 1-jet channels. The SM background (blue) is compared against signal hypotheses for $m_H=200$ GeV and $450$GeV (red and green). The clear separation at $p \rightarrow 1$ validates the network's discriminative power.}
        \label{fig:nnclassscore}
\end{figure}

The AUC improves with the signal mass 
in both channels, reflecting the increasingly distinct kinematics of 
heavier signal events relative to the SM backgrounds. The high AUC 
values at $m_H = 450$~GeV indicate that the network is particularly 
effective at separating high-mass signal from background, while the 
lower AUC at $m_H = 200$~GeV is consistent with the greater kinematic overlap between a 200~GeV resonance and the $WW$ and $t\bar{t}$ backgrounds at low mass.
As shown in Fig.~\ref{fig:nnclassscore}, the DNN successfully isolates the signal in the high-score region, particularly for heavier mass hypotheses where the boosted kinematics are more distinct. The background concentrates near $p \approx 0$ while the signal populates the high-score 
region, confirming that the network has learned to discriminate the two 
hypotheses. 

%%%%%%%%%%%%%%%%%%%%%%%%%%%%%%%%%%%%%%%%%%%%%%%%%

\subsection{Results}
\label{sec:res}

\begin{figure}
    \centering
    % --- First row ---
    \begin{subfigure}[b]{0.45\textwidth}
        \centering
        \includegraphics[width=\textwidth]{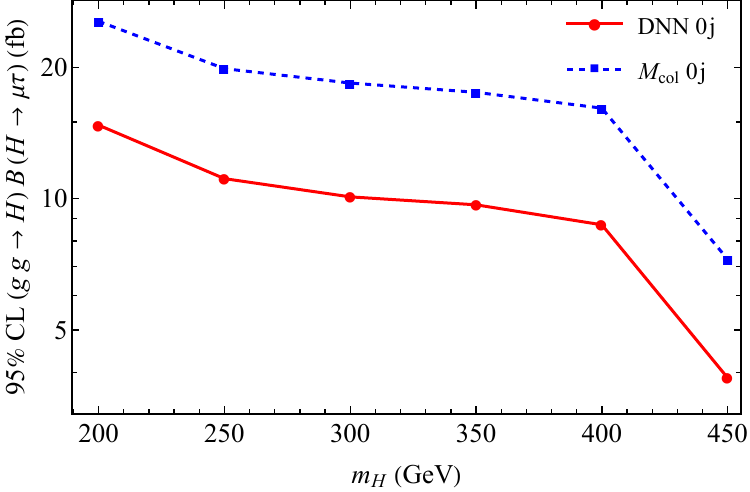}
        \caption{Cross section upper limit for 0-jet}
        \label{fig:xslimit0j}
    \end{subfigure}
    \hfill
    \begin{subfigure}[b]{0.45\textwidth}
        \centering
        \includegraphics[width=\textwidth]{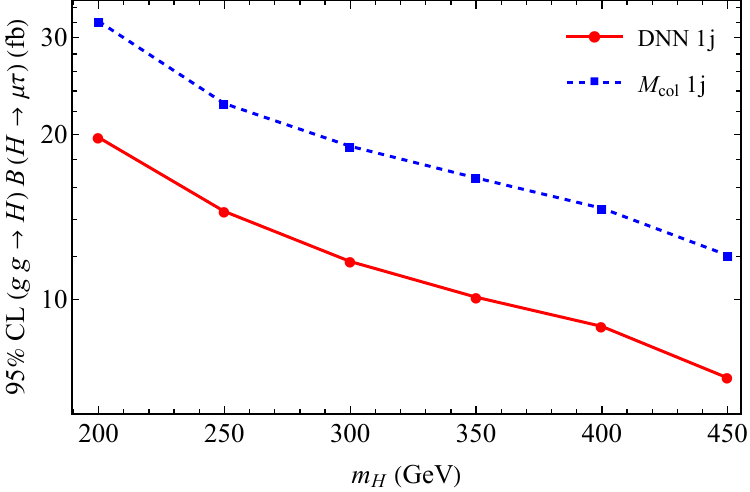}
        \caption{Cross section upper limit for 1-jet}
        \label{fig:xslimit1j}
    \end{subfigure}

    \vspace{0.6cm}

    % --- Second row (centered) ---
    \begin{subfigure}[b]{0.45\textwidth}
        \centering
        \includegraphics[width=\textwidth]{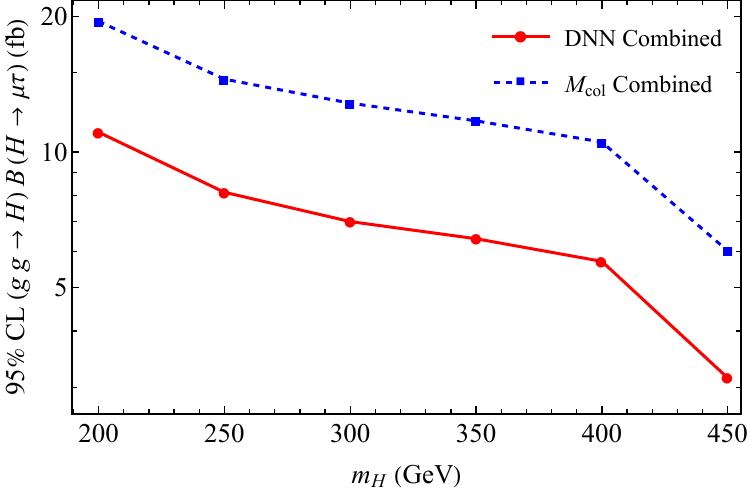}
        \caption{Cross section upper limit, combined channels}
        \label{fig:xslimitcomb}
    \end{subfigure}

    \caption{Expected 95\,\% CL upper limits on the signal cross-section $\sigma(pp \rightarrow H \rightarrow \mu \tau)$ for the (a) 0-jet, (b) 1-jet, and (c) combined channels. The DNN-based limits (red) are compared against the standard $M_{\mathrm{col}}$ baseline (blue).}
    \label{fig:NNvsmcol}
\end{figure}

The expected 95\% CL upper limits on the $H \rightarrow \mu \tau$ production cross-section for the DNN-based and $M_\mathrm{col}$ baseline methods are presented in Table~\ref{tab:limits_comparison} and visualized in Fig.~\ref{fig:NNvsmcol}.

In the 0-jet channel, the $M_\mathrm{col}$ baseline yields limits ranging from 25.50~fb at $m_H=200$~GeV to 7.26~fb at $m_H =450$~GeV. The DNN-based selection provides a significant enhancement in sensitivity across the entire mass range, achieving limits of 14.74~fb and 3.88~fb at the same mass points. This represents a 42--46\,\% reduction in the expected upper limits. The sensitivity gain remains stable at intermediate mass points, scaling positively with the signal mass. This trend correlates with the increased boosted topology and higher AUC values at higher $m_H$. For the 1-jet channel, the $M_\mathrm{col}$ baseline limits are generally weaker than the 0-jet counterparts at low mass but converge at the high-mass boundary. The DNN selection yields a 36--40\,\% improvement over the baseline. The slightly lower relative gain in the 1-jet channel suggests that additional kinematic complexities partially moderate the classifier's performance.

The optimal score threshold ($p_{\mathrm{cut}}$) varies with the signal mass hypothesis. In the 0-jet channel, $p_{\mathrm{cut}}$ rises from 0.54 at $m_H=200$~GeV to a range of 0.82--0.89 for $m_H \in [300, 450]$~GeV. Similarly, in the 1-jet channel, $p_{\mathrm{cut}}$ rises from 0.53 at $m_H=200$~GeV to a plateau of 0.71--0.81 for $m_H \in [250, 450]$~GeV. This behavior reflects the stringent purity requirements optimal for signals that are kinematically well-separated from the background. The combined channel limits (Fig.~\ref{fig:xslimitcomb}) demonstrate that the DNN-driven improvements are robust across event categories, consistently outperforming the $M_\mathrm{col}$ combination.

\begin{table}[htbp]
\caption{Expected 95\% confidence level upper limits on the signal cross section (fb) obtained using the $M_{\mathrm{col}}$ baseline method and the DNN-based analysis in the 0-jet and 1-jet categories.}
\label{tab:limits_comparison}
\begin{ruledtabular}
\begin{tabular}{c ccc ccc}
$m_H$ (GeV) 
& \multicolumn{3}{c}{0-jet category} 
& \multicolumn{3}{c}{1-jet category} \\
\cline{2-4} \cline{5-7}
& $M_{\mathrm{col}}$ & $\quad{p}_{\mathrm{cut}}$ & DNN 
& $M_{\mathrm{col}}$ & $\quad{p}_{\mathrm{cut}}$ & DNN \\
\hline
200 & 25.50 & 0.54 & 14.74 & 32.12 & 0.53 & 19.74 \\
250 & 19.86 & 0.78 & 11.11 & 22.75 & 0.71 & 14.48 \\
300 & 18.41 & 0.85 & 10.10 & 19.03 & 0.77 & 11.74 \\
350 & 17.53 & 0.86 & 9.68  & 16.66 & 0.81 & 10.11 \\
400 & 16.12 & 0.89 & 8.71  & 14.67 & 0.78 & 8.94  \\
450 & 7.26  & 0.82 & 3.88  & 12.04 & 0.79 & 7.20  \\
\end{tabular}
\end{ruledtabular}
\end{table}

\subsection{Importance Analysis of Kinematic Variables}
\label{sec:SHAP}
\begin{figure}
     \centering
     \begin{subfigure}[b]{0.48\textwidth}
         \centering
         \includegraphics[width=\textwidth]{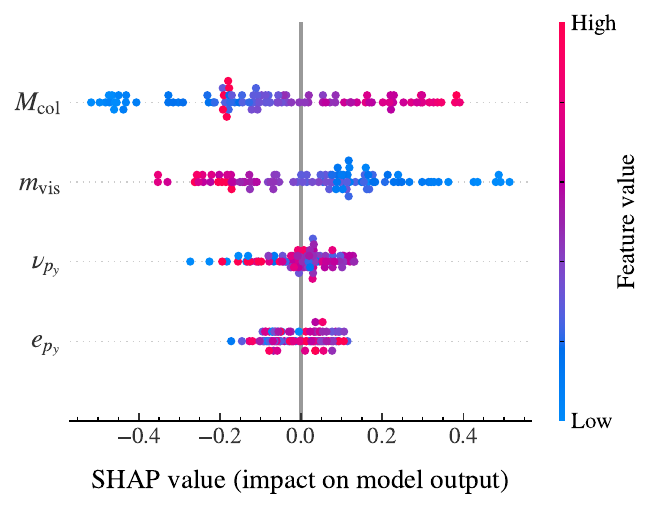}
         \caption{SHAP Importance score for 0 jet channel}
         \label{fig:shap0j}
     \end{subfigure}
     \hfill
     \begin{subfigure}[b]{0.48\textwidth}
         \centering
         \includegraphics[width=\textwidth]{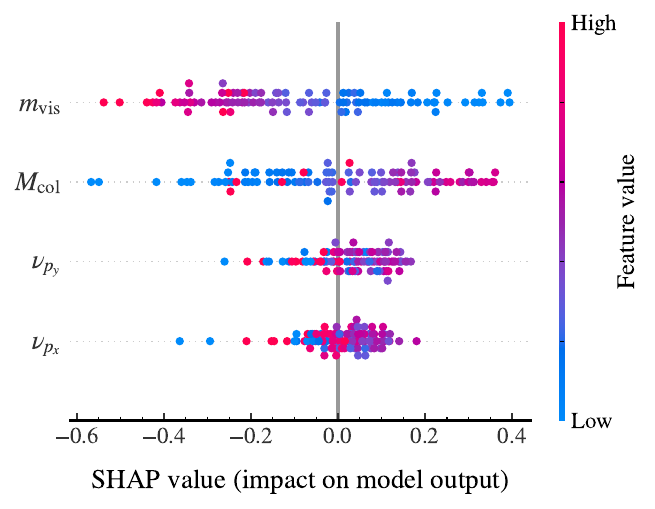}
         \caption{SHAP Importance score for 1 jet channel}
         \label{fig:shap1j}
     \end{subfigure}
\caption{SHAP summary plot for the Neural Network classifier. Four most important features are ranked by their mean absolute SHAP value, representing their global importance. The color scale indicates the feature value (red for high, blue for low), while the x-axis represents the impact on the model output.}
        \label{fig:SHAP Score}
\end{figure}

To interpret the classifier's decision logic, we apply SHAP 
(SHapley Additive exPlanations)~\cite{Lundberg:2017uca} to quantify the 
contribution of each input feature to the classifier output. As illustrated in Fig.\ref{fig:SHAP Score}, where we show four most important parameters based on SHAP analysis, $M_\mathrm{col}$ and $m_\mathrm{vis}$ are the dominant discriminators. While $M_\mathrm{col}$ shows a strong positive correlation with signal probability—consistent with its role as a mass resonance estimator—$m_\mathrm{vis}$ exhibits a significant inverse correlation.

\begin{figure}
     \centering
     \begin{subfigure}[b]{0.48\textwidth}
         \centering
         \includegraphics[width=\textwidth]{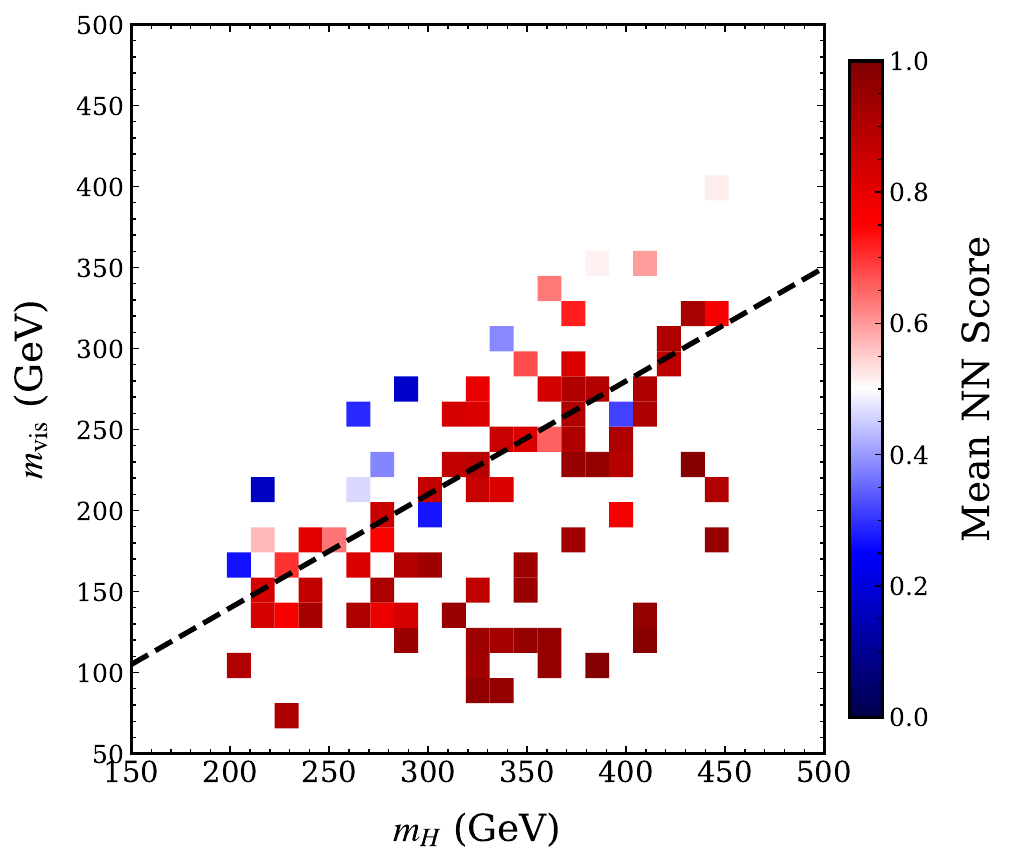}
         \caption{NN score in ($m_H$, $m_\mathrm{vis}$) plane for 0 jet channel}
         \label{fig:nnclassmHmvis0j}
     \end{subfigure}
     \hfill
     \begin{subfigure}[b]{0.48\textwidth}
         \centering
         \includegraphics[width=\textwidth]{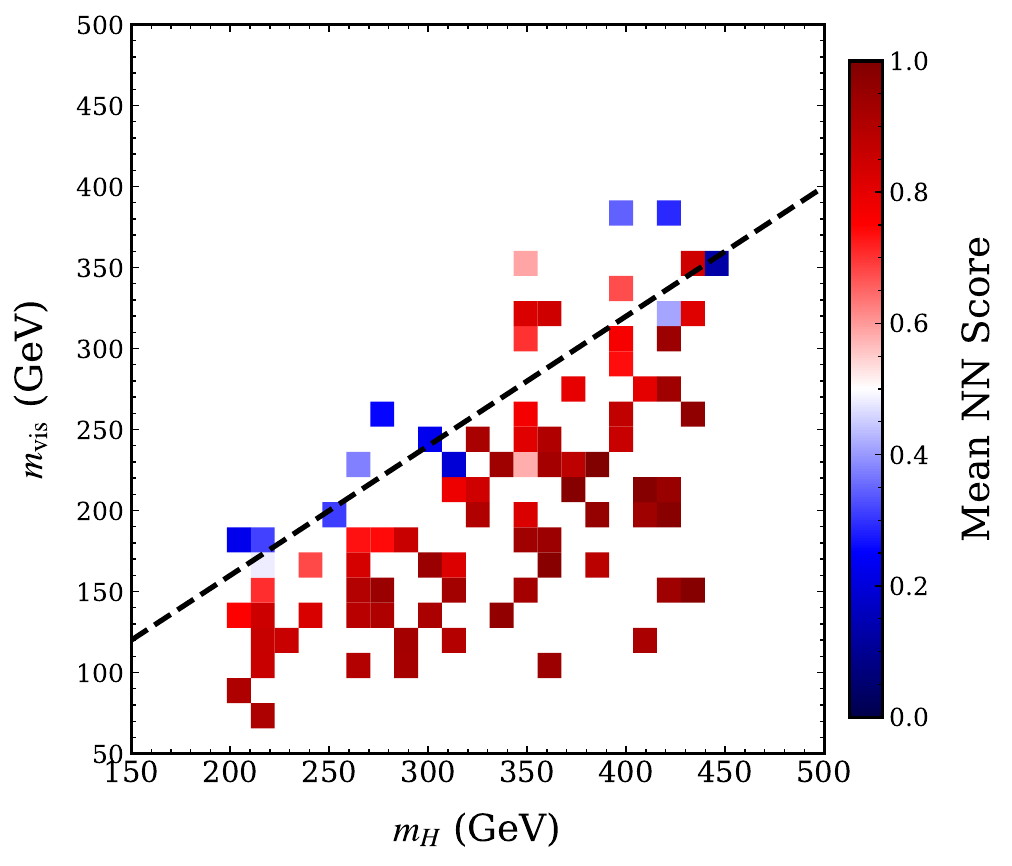}
         \caption{NN score in ($m_H$, $m_\mathrm{vis}$) plane for 1 jet channel}
         \label{fig:nnclassmHmvis1j}
     \end{subfigure}
\caption{Mean DNN classifier score projected onto the ($m_H$, $m_\mathrm{vis}$) plane for the (a) 0-jet and (b) 1-jet channels. The color scale represents the signal-like probability, where the gradient illustrates high signal purity at large $m_H$ and low $m_\mathrm{vis}$. The dashed black lines represent the mass-dependent selection cuts, $m_\mathrm{vis} < 0.7 \cdot m_H$ for the 0-jet channel and $m_\mathrm{vis} < 0.8 \cdot m_H$ for the 1-jet channel. }
        \label{fig:nnclassscoremhmvis}
\end{figure}

The network effectively learns that for an event, a substantial fraction of the tau lepton's momentum is carried by neutrinos, resulting in $m_\mathrm{vis}$ values consistently lower than the true $m_H$. This is visualized in Fig.~\ref{fig:nnclassscoremhmvis}, where the mean DNN score is mapped onto the ($m_H$, $m_\mathrm{vis}$) plane. The resulting gradient demonstrates that the highest signal purity is achieved at large $m_H$  and low $m_\mathrm{vis}$. This observation justifies the use of adding a mass-dependent visible mass cut, $m_\mathrm{vis} < f \cdot m_H$, as a simplified, yet effective alternative to the full multivariate analysis selection.

Motivated by this observation, we impose a simple additional upper bound on the 
visible mass as a fraction $f$ of the signal mass hypothesis, where $f \in [0.3, 1.0]$ is scanned in steps of 0.1. No DNN selection 
is applied; all events passing the visible mass requirement are 
histogrammed directly in $M_\mathrm{col}$. The expected 95\% CL upper 
limit is evaluated using the same test statistic as in 
Eq.~\eqref{eq:likelihood}, and the value of $f$ minimising the expected 
limit is adopted for each mass hypothesis.

%As established in Sec.~\ref{sec:SHAP}, the SHAP analysis identifies $m_\mathrm{vis}$  as the primary discriminating feature alongside $M_{\mathrm{col}}$. The concentration of high DNN scores in the large $m_H$ and low $m_\mathrm{vis}$  region (Fig.~\ref{fig:nnclassscoremhmvis}) indicates that the network prioritizes an upper bound on visible mass. To validate this, we implement a simplified selection: $m_\mathrm{vis}<f\cdot m_H$, without DNN intervention. We find the optimal scaling factor to be $f=0.7$ for the 0-jet channel and $f=0.8$ for the 1-jet channel. 
%The shift in $f$ is attributed to the additional hadronic recoil in 1-jet events, which broadens the $m_\mathrm{vis}$  distribution.

\begin{figure}
    \centering
    % --- First row ---
    \begin{subfigure}[b]{0.45\textwidth}
        \centering
        \includegraphics[width=\textwidth]{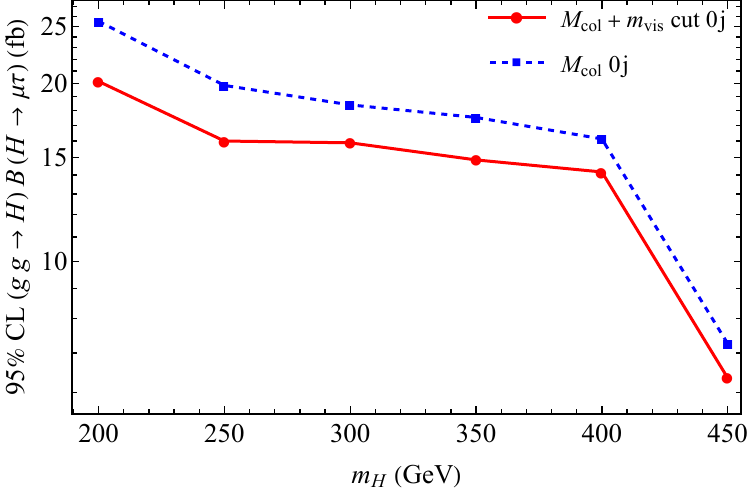}
        \caption{Cross section upper limit for 0-jet, with additional cut on $m_\mathrm{vis}$}
        \label{fig:xslimit0jwmvis}
    \end{subfigure}
    \hfill
    \begin{subfigure}[b]{0.45\textwidth}
        \centering
        \includegraphics[width=\textwidth]{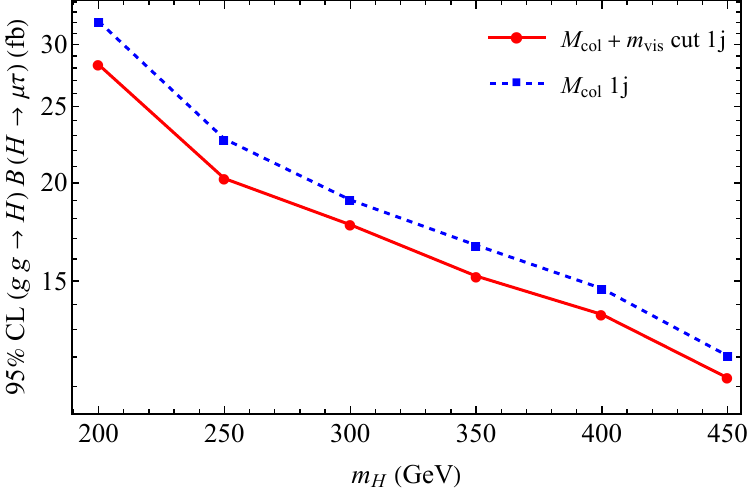}
        \caption{Cross section upper limit for 1-jet, with additional cut on $m_\mathrm{vis}$}
        \label{fig:xslimit1jwmvis}
    \end{subfigure}

    \vspace{0.6cm}

    % --- Second row (centered) ---
    \begin{subfigure}[b]{0.45\textwidth}
        \centering
        \includegraphics[width=\textwidth]{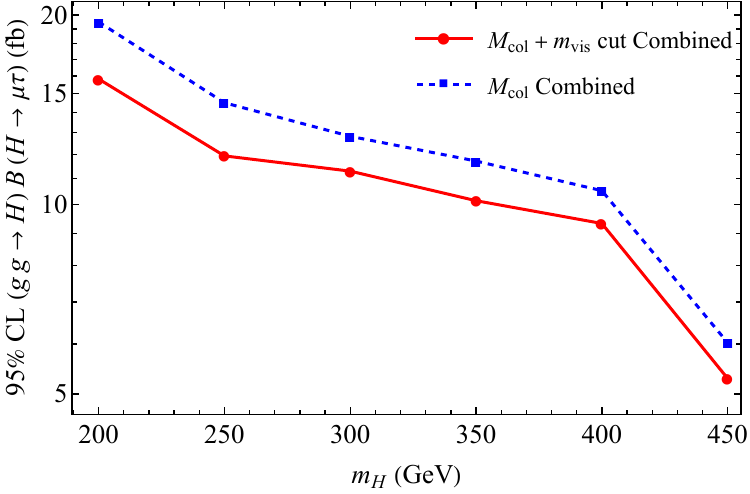}
        \caption{Cross section limit, combined channels, with additional cut on $m_\mathrm{vis}$}
        \label{fig:xslimitcombwmvis}
    \end{subfigure}

    \caption{Comparison of expected 95\,\% CL upper limits on the signal cross-section $\sigma(pp \rightarrow H \rightarrow \mu \tau)$  using the additional mass-dependent $m_\mathrm{vis}$ cut (red) against the $M_\mathrm{col}$ baseline (blue). The $m_\mathrm{vis}$  requirement significantly enhances the signal-to-background ratio by exploiting the neutrino momentum fraction characteristic of the $\tau$ decay.}
    \label{fig:mviscutsaddvsmcol}
\end{figure}

The resulting limits (Fig.~\ref{fig:mviscutsaddvsmcol}) show that this single requirement captures a substantial fraction of the DNN's sensitivity gain. This result is significant, it demonstrates that the improvement is not merely an algorithmic artifact, but a reflection of a physical kinematic property. Specifically, that the LFV signal consistently exhibits a lower $m_\mathrm{vis}$ /$m_H$ ratio than the dominant SM backgrounds. While the full DNN remains superior due to its use of the complete ten-dimensional feature space, the $m_\mathrm{vis}$  cut addition offers a highly interpretable and easily implementable refinement to the standard $M_{\mathrm{col}}$ search strategy.

A direct comparison of $M_\mathrm{col}$ baseline, the $M_\mathrm{col}$ analysis incorporating 
an additional $m_\mathrm{vis}$ cut, and DNN shows a clear hierarchy in sensitivity 
across the full mass range. The DNN achieves the strongest limits in 
all cases, followed by the $M_\mathrm{col}$ analysis with additional $m_\mathrm{vis}$ cut, while the  
$M_\mathrm{col}$ baseline has the lowest sensitivity. The gap between the DNN and 
the $m_\mathrm{vis}$ augmented analysis reflects the additional discriminating 
information captured by the full ten-dimensional input space of the 
network, beyond what is accessible through $m_\mathrm{vis}$ alone. 
Nonetheless, the fact that gain observed by simply adding an $m_\mathrm{vis}$ cut 
to the $M_\mathrm{col}$ baseline remains a significant result. It identifies a concrete and interpretable improvement to 
the baseline analysis strategy that can be implemented without machine 
learning infrastructure.

%%%%%%%%%%%%%%%%%%%%%%%%%%%%%%%%%%%%%%%%%
\section{Mass Reconstruction with Deep Regression}
\label{sec:masspredict}

As discussed in Sec.~\ref{sec:eventselect}, the collinear approximation 
 $M_\mathrm{col}$ carries an inherent systematic bias arising from the sensitivity of the reconstructed momentum fractions to $E_T^\mathrm{miss}$ resolution. This limits the $M_\mathrm{col}$ accuracy and broadens the reconstructed signal resonance. In order to mitigate this limitation, we develop 
a DNN regression model that learns a data-driven correction to the collinear 
mass estimate. The regressor takes the same ten input features as the 
classifier described in Sec.~\ref{sec:NeurNet}. Rather than predicting $m_H$ 
directly, the network is trained to predict the ratio $R = m_H/M_\mathrm{col}$, 
which is of order unity across the entire mass range considered and removes the 
need to learn the overall mass scale from the raw momenta. The final mass 
estimate is then recovered as $m_\mathrm{pred} = R \cdot M_\mathrm{col}$.

For the mass reconstruction task, we use the same high-density signal dataset 
developed for the DNN classifier, which provides a quasi-continuous 
representation of the signal across the 200--450~GeV mass range. This allows 
the network to perform robust mass interpolation and helps prevent overtraining 
on specific mass hypotheses. We evaluate the model's generalization on a 
separate, independent test set consisting of 10 discrete mass points spaced 
between 200 and 450~GeV. In order to ensure that the performance metrics are 
not dominated by statistical noise, we generate $10^4$ events for each of 
these test points.

\begin{figure}
    \centering
    \includegraphics[width=1\textwidth]{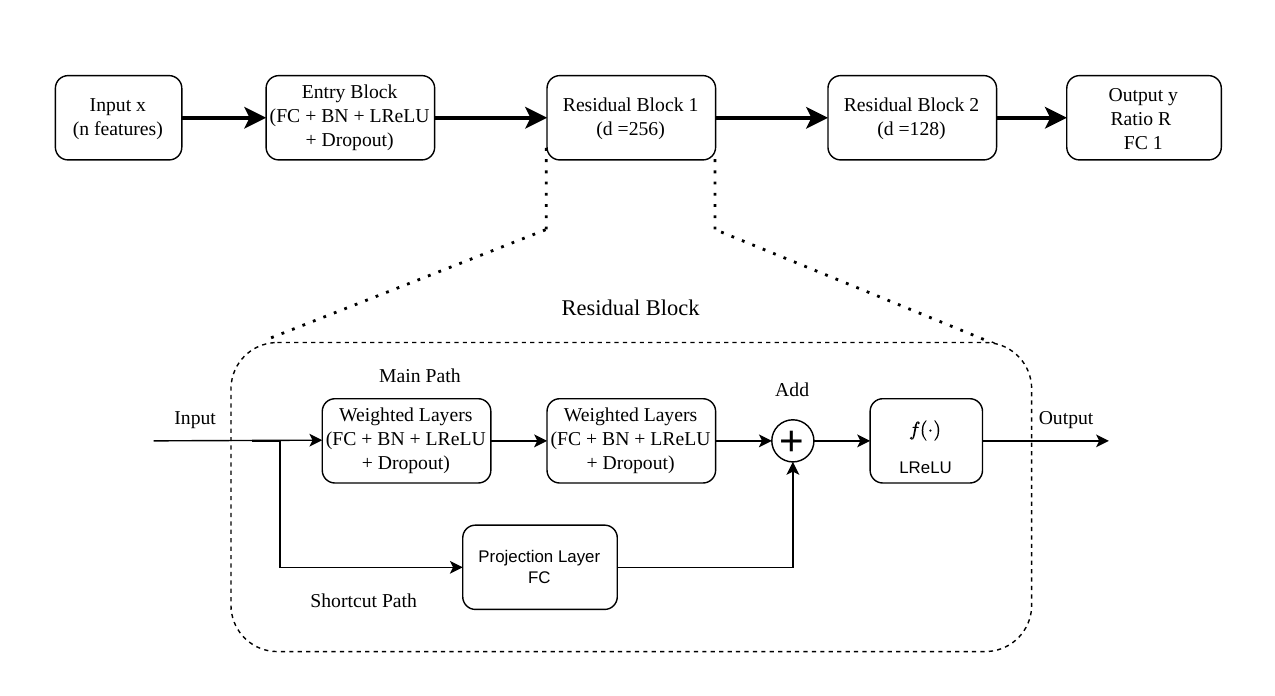}
    \caption{Schematic of the residual network architecture used for Higgs 
    mass regression. The network consists of an entry block followed by two 
    residual units, each with additive skip connections ($\oplus$). The 
    activation function $f(\cdot)$ denotes LeakyReLU ($\alpha = 0.1$), 
    applied after each element-wise addition of the shortcut and residual 
    paths.}
    \label{fig:dnndiagram}
\end{figure}

The mass reconstruction model follows the residual architecture described in 
Fig.~\ref{fig:dnndiagram} and Table~\ref{tab:arch_regression}. We use additive 
skip connections to allow the network to learn perturbative corrections to the 
collinear mass while maintaining stable gradient flow during training. In order 
to handle misreconstructed events in the tails of the $M_\mathrm{col}$ 
distribution, the model is optimized using a Huber loss function 
($\delta = 0.1$), which applies a linear penalty to large residuals. This 
choice, combined with Batch Normalization and LeakyReLU activations, ensures 
consistent performance and numerical stability across the entire 200--450~GeV 
mass range.

Training is performed using an early stopping criterion with a patience 
window of 50 epochs and a dynamic learning rate reduction on plateau; 
convergence is typically achieved within 150 epochs. To ensure stability 
against random weight initialization and avoid local minima, the training 
process is repeated across multiple iterations. The optimal model is 
subsequently selected using a custom weighted score 
$S = \sigma_{\mathrm{SD}} + \lambda\,|\mu - m_H|$, where $\sigma_{\mathrm{SD}}$ 
is the standard deviation of the predicted mass distribution and 
$|\mu - m_H|$ represents the prediction bias. We set $\lambda = 50$ to 
prioritize the minimization of the prediction error over the absolute 
resolution. The full training configuration is summarized in 
Table~\ref{tab:train_regression}.

\begin{table}[h]
\centering
\caption{DNN architecture for mass regression. The model employs residual 
blocks with Batch Normalization (BN) and LeakyReLU activations. Residual 
connections ($\oplus$) denote element-wise addition of the shortcut and 
processed branches.}
\label{tab:arch_regression}
\begin{tabular}{@{}llccc@{}}
\toprule
Block & Layer Type & Units/Dim. & Activation & Dropout \\ \midrule
Input     & Feature Vector     & 10  & ---       & ---  \\ \addlinespace
Entry     & Dense + BN         & 256 & LeakyReLU & 0.2  \\ \addlinespace
Residual 1 & Dense + BN (2$\times$) & 256 & LeakyReLU & 0.2 \\
           & Shortcut $\oplus$  & 256 & LeakyReLU & ---  \\ \addlinespace
Residual 2 & Dense + BN (2$\times$) & 128 & LeakyReLU & 0.1 \\
           & Shortcut $\oplus$  & 128 & LeakyReLU & ---  \\ \addlinespace
Output    & Fully Connected    & 1   & Linear    & ---  \\ \bottomrule
\end{tabular}
\end{table}

\begin{table}[h]
\centering
\caption{Training configuration for the mass regression model.}
\label{tab:train_regression}
\begin{tabular}{ll}
\hline\hline
Hyperparameter    & Value \\
\hline
Optimizer         & Adam \\
Learning rate     & $10^{-3}$, min $10^{-6}$ \\
LR scheduler      & ReduceLROnPlateau (factor 0.5, patience 15) \\
Loss function     & Huber ($\delta = 0.1$) \\
Metric            & Mean Absolute Error (MAE) \\
Batch size        & 1024 \\
Early stopping    & patience 50, monitor val.\ loss \\
Train/test split  & 80\% / 20\% \\
Feature scaling   & StandardScaler (zero mean, unit variance) \\
Target variable   & $m_H / M_\mathrm{col}$ (ratio $\approx 1$) \\
Input features    & $p_x^{\mu}, p_y^{\mu}, p_z^{\mu}, p_x^{e}, p_y^{e},
                    p_z^{e}, p_x^{\nu}, p_y^{\nu}, 
                    m_\mathrm{vis}, M_\mathrm{col}$ \\
\hline\hline
\end{tabular}
\end{table}

\subsection{Reconstruction performance}

\begin{figure}[htbp]
     \centering
     \begin{subfigure}[b]{0.48\textwidth}
         \centering
         \includegraphics[width=\textwidth]{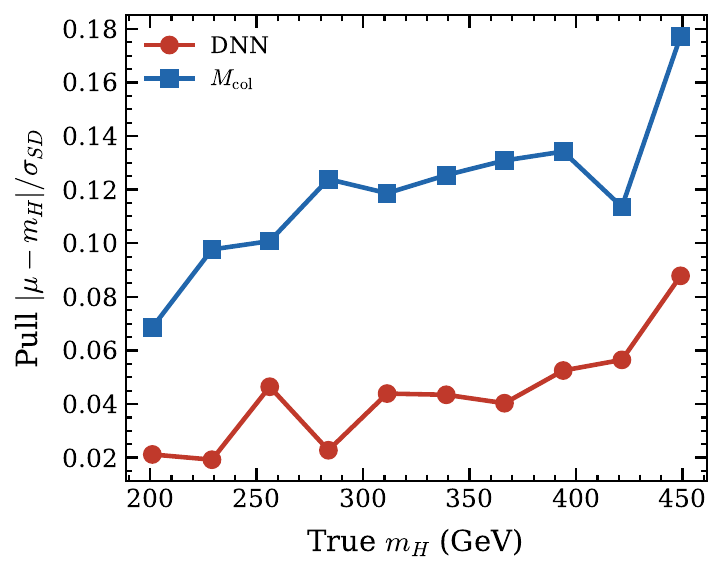}
         \caption{Pull, 0-jet}
         \label{fig:masspred_pull_0j}
     \end{subfigure}
     \hfill
     \begin{subfigure}[b]{0.48\textwidth}
         \centering
         \includegraphics[width=\textwidth]{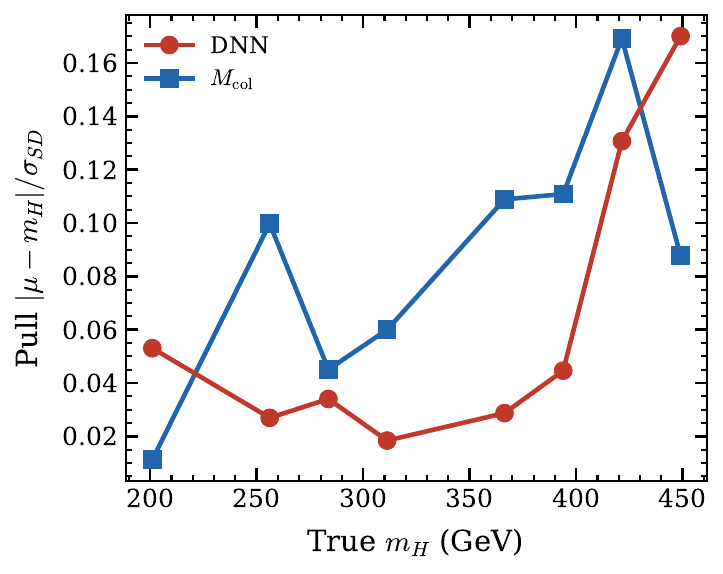}
         \caption{Pull, 1-jet}
         \label{fig:masspred_pull_1j}
     \end{subfigure}
     
     \vspace{1em}
     
     \begin{subfigure}[b]{0.48\textwidth}
         \centering
         \includegraphics[width=\textwidth]{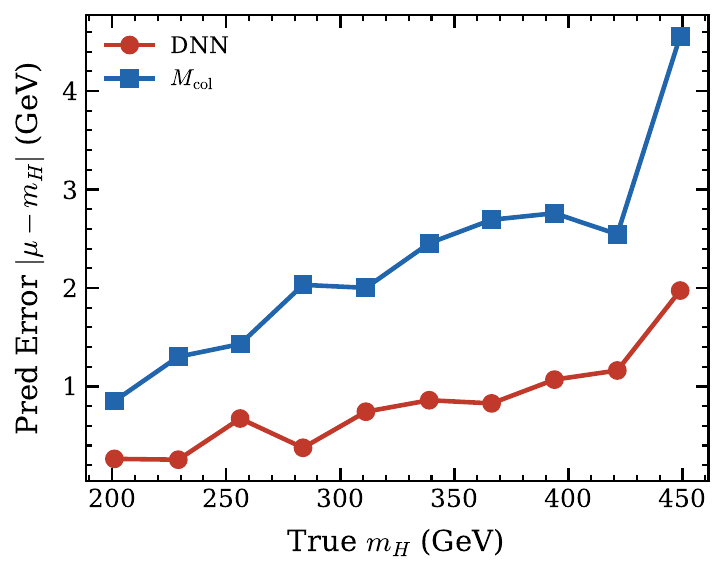} 
         \caption{Prediction error, 0-jet}
         \label{fig:masspred_error_0j}
     \end{subfigure}
     \hfill
     \begin{subfigure}[b]{0.48\textwidth}
         \centering
         \includegraphics[width=\textwidth]{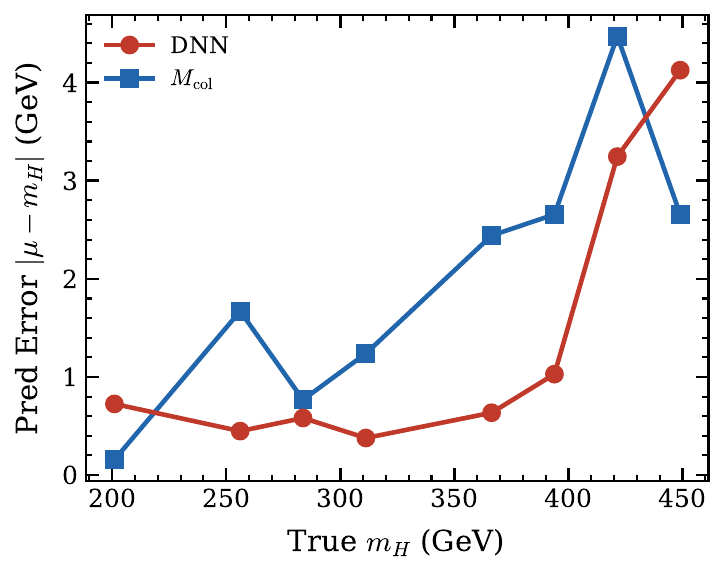} 
         \caption{Prediction error, 1-jet}
         \label{fig:masspred_error_1j}
     \end{subfigure}
     
     \vspace{1em}
     
     \begin{subfigure}[b]{0.48\textwidth}
         \centering
         \includegraphics[width=\textwidth]{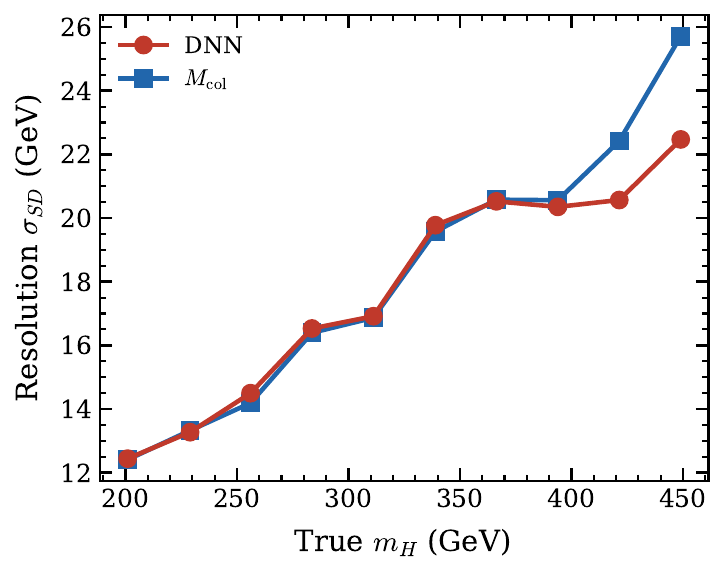} 
         \caption{Resolution, 0-jet}
         \label{fig:masspred_res_0j}
     \end{subfigure}
     \hfill
     \begin{subfigure}[b]{0.48\textwidth}
         \centering
         \includegraphics[width=\textwidth]{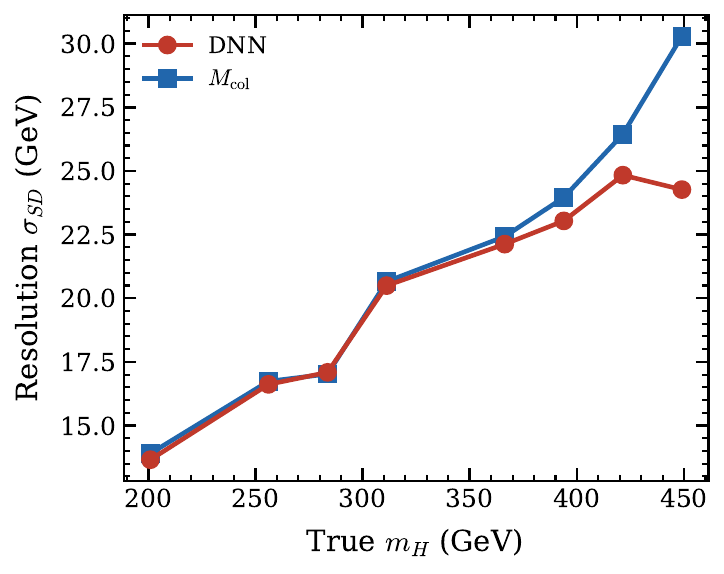} 
         \caption{Resolution, 1-jet}
         \label{fig:masspred_res_1j}
     \end{subfigure}
     \caption{Mass reconstruction performance of the DNN regressor compared 
     to $M_\mathrm{col}$. The normalized pull $(\mu - m_H)/\sigma_{SD}$ 
     (top row), prediction error $|\mu - m_H|$ (middle row), and resolution 
     $\sigma_{SD}$ (bottom row) are shown as a function of the true signal 
     mass $m_H$ for the 0-jet (left) and 1-jet (right) channels.}
     \label{fig:masspred_complete}
\end{figure}

The reconstruction performance is evaluated using three metrics: the 
prediction error $|\mu - m_H|$, the resolution $\sigma_{SD}$, and the 
normalized pull $|\mu - m_H|/\sigma_{SD}$. The results are shown as a 
function of $m_H$ in Fig.~\ref{fig:masspred_complete}.

In the 0-jet channel, the DNN consistently outperforms $M_\mathrm{col}$ in 
terms of prediction error, maintaining an error below 1~GeV for mass points 
up to 400~GeV, compared to approximately 4.6~GeV for $M_\mathrm{col}$ at the 
same mass point. This confirms that the network successfully learns the 
kinematic correlations required to correct the systematic bias
inherent in the collinear approximation, from the $E_T^\mathrm{miss}$ reconstruction effects. In the 1-jet channel, a similar 
improvement is observed up to 350~GeV; the crossover at 450~GeV is attributed 
to boundary effects and limited training statistics at the kinematic edge of 
the sample.

In terms of resolution, both methods exhibit a characteristic degradation as 
$m_H$ increases, driven by the scaling of $E_T^\mathrm{miss}$ and the 
final-state momenta. However, at the upper boundary ($m_H = 450$~GeV), the 
DNN provides a relative resolution improvement of 12\% (0-jet) and 21\% 
(1-jet) over $M_\mathrm{col}$. This suggests that the DNN makes better use 
of the available kinematic information in the 1-jet category to constrain 
the neutrino system.

The normalized pull provides the most rigorous metric of reconstruction 
quality. In the 0-jet channel, the DNN maintains a stable pull below 0.09 
across the entire mass range, while the $M_\mathrm{col}$ pull more than 
doubles toward the high-mass tail, reaching approximately 0.18. In the 1-jet 
channel, the DNN pull exhibits non-monotonic behavior, which reflects the 
varying applicability of the collinear assumption across different boost 
regimes. The notable exception to the overall improvement is the prediction 
error in the 1-jet channel at 200 and 450~GeV, where limited training 
statistics and edge effects cause the DNN to underperform $M_\mathrm{col}$; 
this is a known limitation and a direction for future improvement~\cite{Baldi:2016fzo}.

Overall, the DNN regression provides a consistent improvement in mass 
reconstruction quality relative to the collinear approximation across most 
of the mass range considered. The most significant gain is in prediction 
error: the DNN corrects the systematic overestimation of $M_\mathrm{col}$ at 
most of considered mass range , maintaining errors below 1~GeV up to 400~GeV in the 0-jet channel, 
compared to errors exceeding 2~GeV for $M_\mathrm{col}$ at the same mass 
points. The normalized pull confirms that the DNN achieves a more accurate 
and stable reconstruction across the mass spectrum, with the most marked 
improvement in the 1-jet channel where the collinear approximation is most 
challenged.

%%%%%%%%%%%%%%%%%%%%%%%%%%%%%%%%%%%%%%%%%
\section{Conclusion}
\label{sec:conc}

In this work, we have investigated the sensitivity to lepton-flavor-violating (LFV) Higgs decays, $H \to \mu\tau_e$, within the framework of a Type-III Two-Higgs-Doublet Model. By recasting the 13~TeV CMS search with 35.9~fb$^{-1}$ of integrated luminosity using fast detector simulation, we have demonstrated that modern machine learning techniques can significantly enhance the discovery potential for heavy LFV Higgs bosons in the 200–450 GeV mass range. Our analysis yields three primary conclusions:

First, we established that a DNN-based classifier provides a substantial gain in sensitivity compared to the traditional collinear mass ($M_\mathrm{col}$) selection. By optimizing the score threshold independently for each mass hypothesis, we achieved a 36--46\% reduction in the expected 95\% CL upper limits on the signal cross-section. This improvement is robust across both 0-jet and 1-jet categories, with the most pronounced gains occurring in the high-mass regime where the boosted kinematics of the signal are most distinct from SM backgrounds.

Second, we applied SHAP-based interpretability analysis to interpret the classifier's decisions. This analysis identified the visible mass, $m_\mathrm{vis}$, as one of the main discriminating features. We demonstrate that supplementing the baseline analysis with a simple, mass-dependent cut ($m_\mathrm{vis}<f\cdot m_H$) captures a significant fraction of the DNN’s performance. This result is particularly impactful for experimental implementation, as it provides a physically intuitive and easily verifiable refinement to existing search strategies without requiring a full multivariate infrastructure.

Third, we addressed the intrinsic systematic prediction error of the collinear mass approximation through deep regression. The DNN regressor successfully mitigates the mass prediction bias of $M_\mathrm{col}$ throughout the majority of the explored mass range, maintaining an absolute prediction error below 1 GeV for signals up to 400 GeV. The resulting stability in the normalized pull distributions confirms that the network effectively leverages the full event topology to recover information lost to the collinear assumption.

While the present analysis relies on fast detector simulation and the 
leptonic $\tau$ decay mode only, the consistency of the improvements 
across multiple metrics and both jet categories provides a compelling 
case for adopting these techniques in dedicated experimental analyses. 
The analogous decay channel $H \to \tau e$ is expected to benefit from 
the same DNN-based enhancements, and extending this framework to that 
channel is a natural next step. Subsequent iterations of this analysis 
would also benefit from a full treatment of systematic uncertainties, 
the inclusion of NLO QCD corrections, and the extension to the hadronic 
$\tau$ decay mode. The $m_\mathrm{vis}$-based simplified selection 
identified in this work offers in particular a concrete and immediately 
applicable refinement to the standard $M_\mathrm{col}$ strategy for 
future LFV Higgs searches at the LHC.

%%%%%%%%%%%%%%%%%%%%%%%%%%%%%%%%%%%%%%%%%
\begin{acknowledgments}
 A.\,F. was supported by funding from the Center for Higher Education Funding and Assesment (PPAPT), Ministry of Higher Education, Science and Technology of Republic Indonesia, and also from Indonesia Endowment Fund for Education Agency (LPDP) as well as Indonesian Education Scholarship (BPI) under funding ID 202209092458. R.P. was supported by  Direktorat Penelitian dan Pengabdian kepada Masyarakat, Direktorat Jenderal Riset dan Pengembangan, Kementerian Pendidikan Tinggi, Sains dan Teknologi
Republik Indonesia with contract number 7939/LL4/PG/2025; III/LPPM/ 2025-06/154-PE and 125/C3/DT.05.00/PL/2025. F.\,T.\,A. and\, B.\,E.\,G. are supported by PPMI KK FMIPA ITB 2026 with contract number 062/IT1.C02/SK-DA/2026. 

\end{acknowledgments}

\appendix

%%%%%%%%%%%%%%%%%%%%%%%%%%%%%%%%%%%%%%%%%
%\bibliographystyle{unsrt}
\bibliography{paper_v1} 
\bibliographystyle{apsrev4-2}
%%%%%%%%%%%%%%%%%%%%%%%%%%%%%%%%%%%%%%%%%%%%%%%%%%%%%%%%%%%%%%%%%%%%%%%%%%

\end{document}